\documentclass[journal]{IEEEtran}
\usepackage{cite}
\usepackage{amsmath,amssymb,amsfonts}
\usepackage{algorithmic}
\usepackage{graphicx}
\usepackage{textcomp}
\def\BibTeX{{\rm B\kern-.05em{\sc i\kern-.025em b}\kern-.08em
    T\kern-.1667em\lower.7ex\hbox{E}\kern-.125emX}}
\usepackage{color}
\usepackage{bm}
\usepackage{algorithm}
\usepackage{algorithmic}

\ifCLASSOPTIONcompsoc
    \usepackage[caption=false, font=normalsize, labelfont=sf, textfont=sf]{subfig}
\else
\usepackage[caption=false, font=footnotesize]{subfig}
\fi

\usepackage{soul}

\begin{document}

\title{2-D Coherence Factor for Sidelobe and Ghost Suppressions in Radar Imaging}

\author{Shiyong Li, \IEEEmembership{Member, IEEE}, Moeness Amin, \IEEEmembership{Fellow, IEEE}, Qiang An, 
Guoqiang Zhao, 
and Houjun Sun   
        
\thanks{Manuscript received March, 2019. This work was performed while Dr. S. Li was a Visiting Research Scholar in the Center for Advanced Communications, Villanova University. The work of Shiyong Li, Guoqiang Zhao, and Houjun Sun was supported by the National Natural Science Foundation of China under Grant 61771049. }
\thanks{S. Li, G. Zhao and H. Sun are with the Beijing Key Laboratory of Millimeter Wave and Terahertz Technology, Beijing Institute of Technology, Beijing 100081, China. (e-mail: lisy\_98@bit.edu.cn).}
\thanks{M. Amin is with the Center for Advanced Communications, Villanova University, Villanova, PA 19085 USA (e-mail: moeness.amin@villanova.edu).}
\thanks{Q. An is with the department of Biomedical Engineering, Fourth Military Medical University, Xi’an 710032, China (e-mail: qan01@villanova.edu).}
}

\markboth{IEEE}%
{Shiyong Li: 2D Coherence Factor for Sidelobes and Ghosts Suppression in Near-field Radar Imaging}

\maketitle

\begin{abstract}


The coherence factor (CF) is defined as the ratio of coherent power to incoherent power received by the radar aperture. The incoherent power is computed by the multi-antenna receiver based on only the spatial  variable.  In this respect,  it is a one-dimensional (1-D) CF, and thereby the image sidelobes in down-range cannot be effectively suppressed. We propose a two-dimensional (2-D) CF by supplementing the 1-D CF by an incoherent sum dealing with the frequency dimension. In essence, we employ both spatial diversity and frequency diversity which, respectively, enhance imaging quality in cross range and range. Simulations and experimental results are provided to demonstrate the performance advantages of the proposed approach.





\end{abstract}


\IEEEpeerreviewmaketitle

\section{Introduction}

\IEEEPARstart{R}{adar} imaging has wide application areas including through-the-wall imaging \cite{Amin_2010book}, target scattering diagnosis and recognition \cite{knott}, and Earth remote sensing \cite{remote_sensing}.


Most radar imaging approaches are based on the Born approximation, which is a linearized model of the electromagnetic (EM) scattering phenomena \cite{Born}.
Linear models can be solved using robust and computationally efficient algorithms. 
However, linear models describe only direct scattering from targets to radar, and neglect multiple scattering phenomena, such as target-to-target and target-to-environment interactions \cite{multipath_insight}. These multipaths have adverse effects on imaging quality. They produce spurious targets as ``ghosts'' in the image scene, thus, increasing clutter and false alarms \cite{comparative_analysis_amin}.
On the other hand, high sidelobes in the radar image could also be recognized as spurious targets and lead to a severe decline in target detection and classification \cite{sidelobes_access}. 

In this context, and towards improving system performance, many research efforts have been dedicated to multipath suppression and exploitation   \cite{ghosts_exploit_amin,ghosts_mitigation}. To achieve the latter, geometrical information of main reflectors is used to map the ghosts back to the true target position, thereby increasing signal-to-clutter ratio at the target location \cite{multipath_amin}. 
In \cite{multipath_localization}, virtual sensors stemming from specular wall reflections were used to localize an indoor target utilizing only one physical sensor deployment.
Microwave images were reconstructed by combining the multipath information as a prior with compressive sensing in \cite{multipath_cs}. 


This work is focused on suppressing   imaging ghosts and target  sidelobes. 
Approaches exploiting  aspect-dependent characteristics  to suppress multipath ghosts by subaperture imaging strategies were presented in \cite{ghosts_subaperture,ghosts_subaperture1}.
The total variation constrained sparse reconstruction method was presented in \cite{qiang} to mitigate ghosts for detection of human behind walls.
A simple but powerful approach for enhanced radar imaging is the coherence factor (CF)  filtering  that alleviates clutter by suppressing its low-coherence features\cite{awpl_cf_1st,comparative_analysis_amin}.
The CF is defined as the ratio of the coherent power received by the radar aperture to the incoherent power, and it is calculated for every position in the image scene. CF was first applied in ultrasonic imaging \cite{cf_ultrasonic,cf_sidelobe_ultra}, and then successfully employed in through-the-wall radar imaging \cite{awpl_cf_1st}.  

At the target location, CF takes a unit value, signifying that the coherent and non-coherent summations assume equal values. However, considering a location where there is no target, the coherent sum becomes much lower than the non-coherent sum, yielding a small CF value. Therefore, ghosts and sidelobes can be suppressed by the mere multiplication of the radar image with the CF map. The latter comprises the CFs at all locations. In essence, the CF utilizes  angle diversity, rendered by the multi-sensor configuration, to suppress undesired image components.  


However, the incoherent summation of CF is implemented only along the array aperture, or the azimuth dimension, specifically, the phase delays corresponding a presumed target are adjusted prior to their summation. In this respect, the CF can be used to attenuate image components along the cross-range, but has minor effects on the suppression of sidelobes along the range dimension. This calls for employing another diversity in the frequency dimension to deal with this problem. 

In this communication, we expand the incoherent summation in the original CF over the  frequency dimension. 
Since we use two types of diversities namely, angle and frequency, the sidelobes and ghosts of low-coherence features will be better suppressed compared with the original CF. This approach is also applicable to the phase coherence factor (PCF) \cite{pcf1} -- another commonly used method to suppress undesired image components \cite{pcf_sidelobes}. It depends on the standard deviation of the complex exponential function in regard to the phase distribution of multiple antenna positions for an image pixel \cite{sign_cf_qinghuo}. 

The remainder of this communication is as follows. Section II presents the formulation of the near-field radar imaging for a generalized configuration. In Section III, we propose the 2-D coherence factor. Numerical simulations and experimental results are shown in Section IV. Finally, concluding remarks are presented in Section V.

\section{Near-Field Radar Imaging by BP Algorithm}

\begin{figure}[!t]
	\centering
	\includegraphics[width=2.6in]{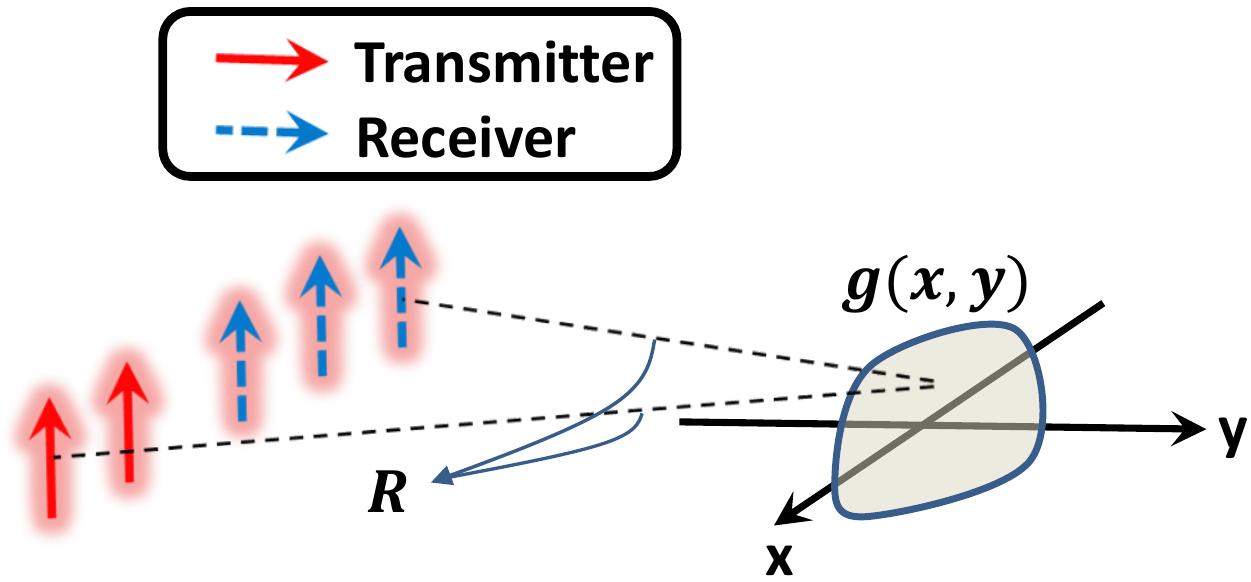}
	\caption{Generalized MIMO imaging geometry.}
	\label{mimo_model}
\end{figure}

We consider a generalized radar imaging geometry with multiple-input multiple-output (MIMO) array, as illustrated in Fig. \ref{mimo_model}. The imaging system has  
$M$ transmitters, which illuminate the target $g(x,y)$, and $N$ receivers. 



Under Born approximation, the scattered electromagnetic (EM) wave is given by,
\begin{equation}\label{scat_wave1}
E(\vec{r}_{\textrm{T}},\vec{r}_{\textrm{R}},f)\!=\!\iint_G g(\vec{r})e^{-\mathrm{j2\pi}f\frac{R(\vec{r},\vec{r}_{\textrm{T}},\vec{r}_{\textrm{R}})}{c}}\mathrm{d}x\mathrm{d}y, 
\end{equation}
where $f$ is the operating frequency, 
$\vec{r}_{\textrm{T}}=(x_{\textrm{T}},y_{\textrm{T}})$ represents the transmitter position, $\vec{r}_{\textrm{R}}=(x_{\textrm{R}},y_{\textrm{R}})$ is the receiver position, 
$\vec{r}=(x,y)$ denotes the coordinates of the imaging scene, $g(\vec{r})$ represents the scattering coefficients of the target, 
{$G$ represents the imaging scene, }
$c$ is the speed of EM wave in free space, and $R(\vec{r},\vec{r}_{\textrm{T}},\vec{r}_{\textrm{R}})$ is the two-way distance from the transmitting antenna to the target to the receiving antenna, 

\begin{align} \label{RR}
&R(\vec{r},\vec{r}_{\textrm{T}},\vec{r}_{\textrm{R}}) \nonumber \\
&=\sqrt{(x-x_{\textrm{T}})^2+(y-y_{\textrm{T}})^2} + \sqrt{(x-x_{\textrm{R}})^2+(y-y_{\textrm{R}})^2}
\end{align}


The forward model in \eqref{scat_wave1} is linear. This means that only the direct reflections from the scattering centers are considered. 
The radar image can be reconstructed by the back-projection (BP) algorithm, as shown by the following two steps:
\begin{equation}\label{bp_1}
y(\vec{r},\vec{r}_{\textrm{T}},\vec{r}_{\textrm{R}})=\int_f E(\vec{r}_{\textrm{T}},\vec{r}_{\textrm{R}},f)e^{\mathrm{j2\pi}f\frac{R(\vec{r},\vec{r}_{\textrm{T}},\vec{r}_{\textrm{R}})}{c}}f\mathrm{d}f
\end{equation}

\begin{equation}\label{bp_2}
g(\vec{r})=\int_{\vec{r}_{\textrm{T}}} \int_{\vec{r}_{\textrm{R}}} y(\vec{r},\vec{r}_{\textrm{T}},\vec{r}_{\textrm{R}}) \mathrm{d}\vec{r}_{\textrm{T}} \mathrm{d}\vec{r}_{\textrm{R}}, 
\end{equation}
Eq. \eqref{bp_1} can be efficiently computed by an inverse fast Fourier transform (IFFT) followed by an interpolation from the uniform grids to the grids corresponding to $R(\vec{r},\vec{r}_{\textrm{T}},\vec{r}_{\textrm{R}})$. 

Transmitted signals are typically designed as step-frequency EM waves. This implies that $f$ is chosen as $f_i=f_0+i\Delta f$, where $f_0$ and $\Delta f$ are, respectively, the initial frequency and frequency step size, and $i=0, 1, \cdots, I-1$.

\section{2-D Coherence Factor}

The coherence factor was first applied in ultrasonic imaging \cite{cf_ultrasonic}, and then utilized in though-the-wall radar imaging \cite{awpl_cf_1st,comparative_analysis_amin} to enhance image quality. CF measures the ratio between the total coherent power to the total incoherent power received by the antenna aperture, and is expressed as \cite{cf_ultrasonic},
\begin{equation} \label{cf_azimu}
\textrm{CF}(\vec{r})=\frac{\left| \sum_{m=0}^{M-1}\sum_{n=0}^{N-1} y_{mn}(\vec{r})\right|^2}{MN\sum_{m=0}^{M-1}\sum_{n=0}^{N-1} \left|y_{mn}(\vec{r})\right|^2}, \qquad \vec{r}\in G 
\end{equation}
where $y_{mn}(\vec{r})$ represents $y(\vec{r},\vec{r}_{\textrm{T}_m},\vec{r}_{\textrm{R}_n})$ in \eqref{bp_1}.
Notice that the numerator in \eqref{cf_azimu} is the image $g(\vec{r})$ in \eqref{bp_2}. 

The values of CF vary from zero to one and provide information about the low- and high-coherence regions in the image scene. The CF enhanced image is obtained by multiplication of the BP image with the relevant CF map, i.e.,
\begin{equation} \label{im_CF_azimu}
g^{\textrm{CF}}(\vec{r})=\textrm{CF}(\vec{r})\cdot g(\vec{r}) 
\end{equation}

According to \eqref{cf_azimu} and \eqref{im_CF_azimu}, features with low coherence, such as ghosts and sidelobes, will be suppressed or significantly attenuated. 

From the perspective of tomography \cite{ct_sar}, the radar image is obtained by coherent summation, which amounts to first correcting the phase delays from the target scattering centers, then summing the corrected EM waves along all  array spatial  channels. These two steps correspond to equations \eqref{bp_1} and \eqref{bp_2}, respectively. 
At the target location, the phase delays induced by wave propagation are fully corrected. Thus, the coherent sum is responsible for providing $MN$ gain at the target angular location which represents the center of the main lobe of the point spread function. At other positions, however, where there are no targets, the different residual phases produce positive and negative terms in the summation, giving rise to ghosts and sidelobes of the point spread function. 


We maintain that the term $y_{mn}(\vec{r})$ represents the downrange profile mapping at a 2-D image scene, i.e., the phase delays induced by the wave propagation are corrected for each transmitting and receiving antenna-pair. 
And the range profile $y_{mn}(\vec{r})$ for the $m$-th transmitter and $n$-th receiver is illustrated by an arc defined by their distance separation. 
At the target location, the incoherent summation over all array elements (antenna pairs) should be equal to the square of the coherent sum, i.e., rendering the corresponding value of CF equal to one. However, for other points along the arc traces, the incoherent sum remains high, while the coherent sum will be reduced. This results in smaller values of CF which would then suppress ghosts and sidelobes upon multiplication of the radar image.  

However, the incoherent sum in the denominator only incorporates  the aperture direction, since according to \eqref{bp_1},  $y_{mn}(\vec{r})$ has already coherently compensated for the phase delays along the frequency variable. 
Thus, we can infer that the CF values in the range dimension will be larger than those in the azimuth dimension.

\begin{figure}[!t]
	\centering
	\includegraphics[width=2.5in]{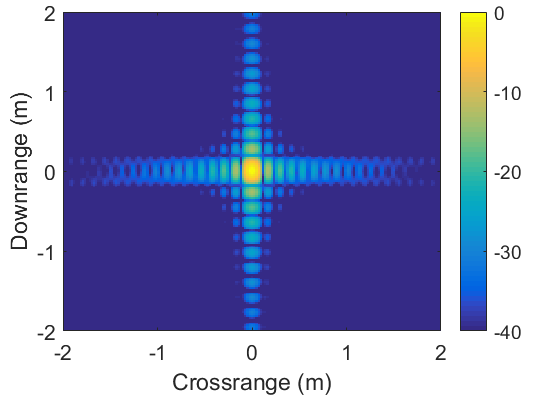}
	\caption{Imaging result of a point target by BP algorithm.}
	\label{bp_im}
\end{figure}

\begin{figure}[!t]
	\centering
	\subfloat[]{\label{a}
		\includegraphics[width=1.69in]{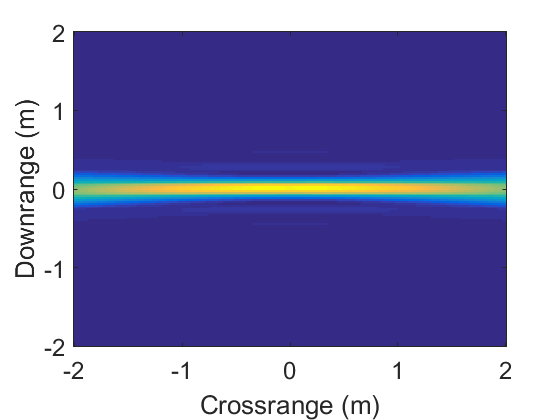}}
	\hfill
	\subfloat[]{\label{b}
		\includegraphics[width=1.69in]{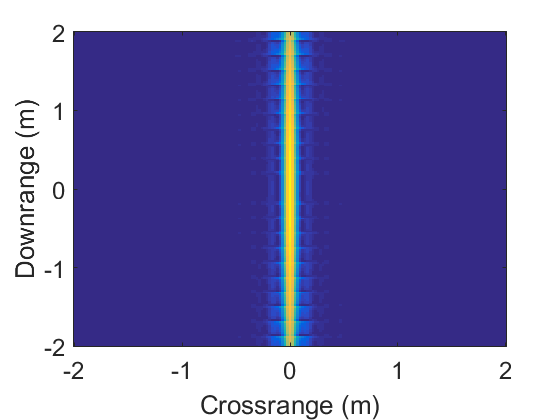}}
	\\

	\caption{(a) The denominator of the original CF, and (b) the related CF map.}
	\label{cf_map_denominator}
\end{figure}

\begin{figure}[!t]
	\centering
	\includegraphics[width=2.5in]{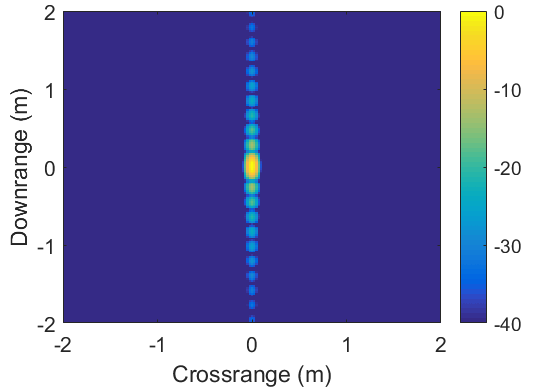}
	\caption{The original CF enhanced image.}
	\label{cf_theta_im}
\end{figure}

We give a simple example to show the CF map of a point target image in Fig. \ref{bp_im}, which represents a turntable inverse synthetic aperture (ISAR) radar imaging result.
The denominator of CF is shown in Fig. \ref{cf_map_denominator}\subref{a}. Clearly, large values are located along the azimuth direction due to the incoherent sum over the aperture. This leads to smaller CF values along the azimuth direction and larger values along the range direction, as illustrated in Fig. \ref{cf_map_denominator}\subref{b}. The enhanced radar image is shown in Fig. \ref{cf_theta_im}, which is consistent with the above analysis. The sidelobes in the range dimension are much higher than those in the azimuth dimension.

To improve image quality in range, we can introduce another incoherent sum over the frequency dimension with the goal of suppressing range sidelobes. To invoke the frequency dependency, we express the BP algorithm as follows, 

\begin{equation}\label{bp_f1}
y(\vec{r},f)=\int_{\vec{r}_{\textrm{T}}} \int_{\vec{r}_{\textrm{R}}} E(\vec{r}_{\textrm{T}},\vec{r}_{\textrm{R}},f)e^{\mathrm{j2\pi}f\frac{R(\vec{r},\vec{r}_{\textrm{T}},\vec{r}_{\textrm{R}})}{c}}\mathrm{d}\vec{r}_{\textrm{T}} \mathrm{d}\vec{r}_{\textrm{R}}
\end{equation}

\begin{equation}\label{bp_f2}
g(\vec{r})=\int_{f} y(\vec{r},f) f\mathrm{d}f, 
\end{equation}

Thus, the CF with incoherent sum over the different frequencies, denoted as $\textrm{CF}^f$, is given by,  
\begin{equation} \label{cf_f}
\textrm{CF}^f(\vec{r})=\frac{\left| \sum_{i=0}^{I-1} y_i(\vec{r})\right|^2}{I\sum_{i=0}^{I-1} \left|y_i(\vec{r})\right|^2}, \qquad \vec{r}\in G 
\end{equation}
where $y_i(\vec{r})$ represents $y(\vec{r};f_i)$ in \eqref{bp_f1} with $f=f_i$. It is noted that the numerators in \eqref{cf_f} and \eqref{cf_azimu} assume equal values since they both represent the same near-field radar images reconstructed by coherent summation after compensating for the phase delays.

\begin{figure}[!t]
	\centering
	\subfloat[]{\label{a}
		\includegraphics[width=1.69in]{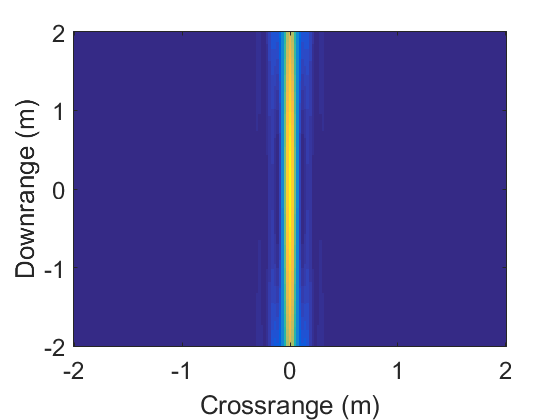}}
	\hfill
	\subfloat[]{\label{b}
		\includegraphics[width=1.69in]{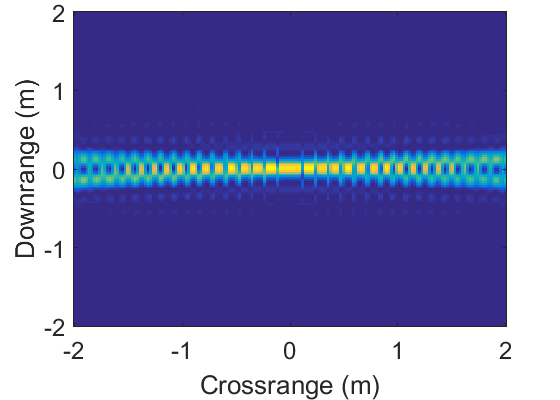}}
	\\
	
	\caption{(a) The denominator of the $\textrm{CF}^f$ with incoherent sum in frequency dimension, and (b) the related $\textrm{CF}^f$ map.}
	\label{cf_f_map_denominator}
\end{figure}

\begin{figure}[!t]
	\centering
	\includegraphics[width=2.5in]{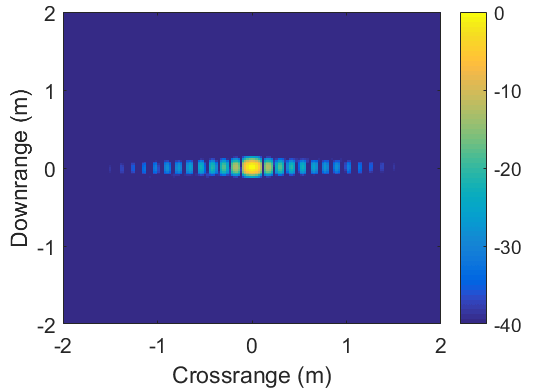}
	\caption{The $\textrm{CF}^{f}$ enhanced image.}
	\label{cf_f_im}
\end{figure}

Note that the incoherent summation in  \eqref{cf_f} is performed in the frequency dimension, enabling effective suppression in the range sidelobes. However, in this case, the sidelobes along the cross-range remain pronounced. In essence, the role of $\textrm{CF}^f$ is opposite to that of CF. 
The $\textrm{CF}^f$ map, its denominator, and the corresponding enhanced radar image are demonstrated in Figs. \ref{cf_f_map_denominator} and \ref{cf_f_im}. 

\begin{figure}[!t]
	\centering
	\includegraphics[width=2.5in]{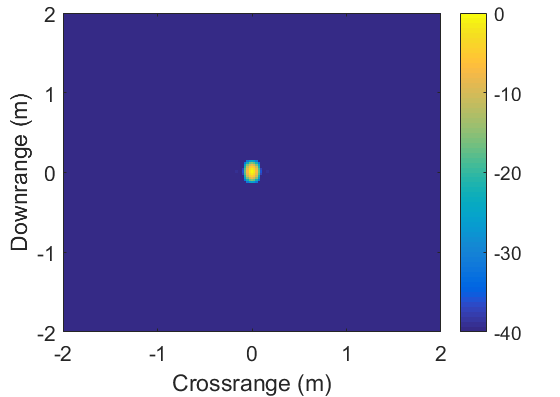}
	\caption{The 2-D $\textrm{CF}$ enhanced image.}
	\label{2d_cf_im}
\end{figure}

In order to reap the benefits of both coherence factors, we introduce the 2-D CF which would effectively suppress  the sidelobes in the azimuth and range dimensions simultaneously. That is, 

\begin{equation} \label{im_CF_2d}
g^{\textrm{CF}}(\vec{r})=\textrm{CF}^{\textrm{2D}}(\vec{r})\cdot g(\vec{r}) 
\end{equation}
where $\textrm{CF}^{\textrm{2D}}(\vec{r})=\textrm{CF}(\vec{r})\cdot \textrm{CF}^{f}(\vec{r})$.

The 2-D CF enhanced image is shown in Fig. \ref{2d_cf_im}. Clearly, the sidelobes along the two dimensions are vividly suppressed. The effect of CF, $\textrm{CF}^{f}$, and $\textrm{CF}^{\textrm{2D}}$ on ghost suppression is analogous to that on the target sidelobes, and will be shown in Section IV. 



Next, we carry the same concept of combined spatial-frequency factor to another commonly  used tool for clutter suppression -- phase coherence factor (PCF) defined as\cite{pcf1},


\begin{align}\label{pcf_theta}
&\textrm{PCF}(\vec{r})=1-\textrm{std}\left(e^{j\angle\mathbf{Y}_{\vec{r}}^{\textrm{azi}}} \right) \nonumber \\
&\angle \mathbf{Y}_{\vec{r}}^{\textrm{azi}}\!=\!\left\{ \angle y_{mn}(\vec{r}), m\!=\!0,1,\cdots,M-1, n\!=\!0,1,\cdots,N-1\right\}
\end{align}
where $\angle y_{mn}(\vec{r})$ denotes the phase of $y_{mn}(\vec{r})$. The standard deviation of the complex exponential term is given by $\textrm{std}\left(e^{j\angle\mathbf{Y}_{\vec{r}}^{\textrm{azi}}}\right)=\sqrt{\textrm{std}^2(\cos\angle{\mathbf{Y}_{\vec{r}}^{\textrm{azi}}})+\textrm{std}^2(\sin\angle{\mathbf{Y}_{\vec{r}}^{\textrm{azi}}})}$.

Note that PCF is defined as a function of the phase dispersion. When $\angle\mathbf{Y}_{\vec{r}}^{\theta}$ distributes uniformly in the range $[-\pi, \pi]$, the standard deviation of the exponential term is unity and $\textrm{PCF}$ reaches zero. Conversely, if all the phases are equal, the standard deviation is zero and $\textrm{PCF}$ becomes unity. In between, $\textrm{PCF}$ takes much smaller values than one for out-of-focus ghosts and sidelobes\cite{pcf1}.

Similar to CF, $\textrm{PCF}$ is also only defined along the aperture direction, and as such, becomes ineffective for range sidelobe suppression. We can define a new PCF by calculating standard deviation in the frequency dimension as,
\begin{align} \label{pcf_f}
\textrm{PCF}^f(\vec{r})&=1-\textrm{std}\left(e^{j\angle\mathbf{Y}_{\vec{r}}^{f}} \right) \nonumber \\
\angle \mathbf{Y}_{\vec{r}}^f&=\left\{ \angle y_i(\vec{r}), i=0,1,\cdots,I-1\right\}
\end{align}
where $y_i(\vec{r})$ denotes the image obtained at the $i\textrm{th}$ frequency described in \eqref{bp_f1}. 

\begin{figure}[!t]
	\centering
	\subfloat[]{
		\includegraphics[width=1.7in]{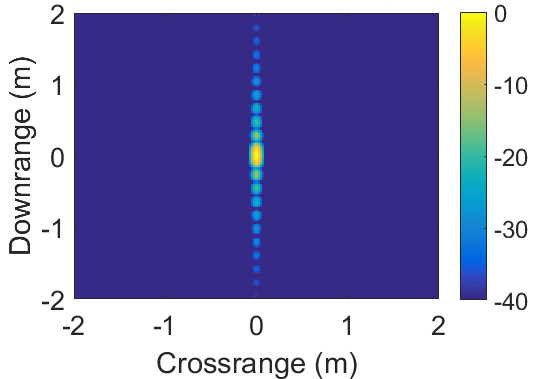}}
	\label{a}\hfill
	\subfloat[]{%
		\includegraphics[width=1.7in]{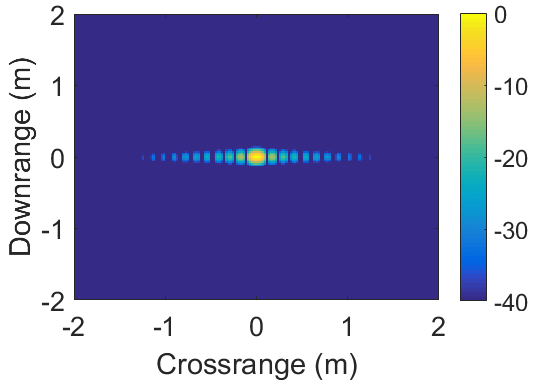}}
	\label{b}\\
	
	\caption{Enhanced images by (a) the original PCF, and (b) the $\textrm{PCF}^f$ with standard deviation in frequency dimension.}
	\label{pcf_theta_f_map}
\end{figure}

\begin{figure}[!t]
	\centering
	\includegraphics[width=2.5in]{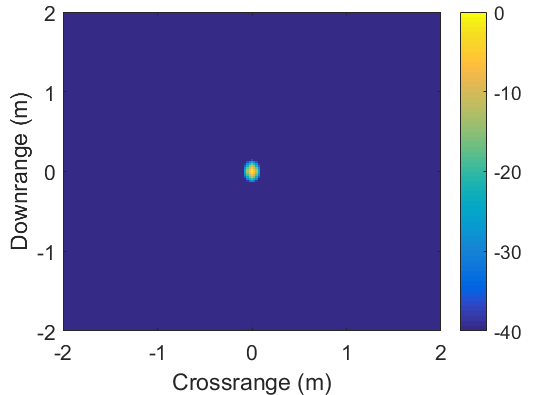}
	\caption{The 2-D $\textrm{PCF}$ enhanced image.}
	\label{2d_pcf_im}
\end{figure}

Thus, the corresponding 2-D PCF can be expressed as,
\begin{equation} \label{pcf2d}
\textrm{PCF}^{\textrm{2D}}(\vec{r})=\textrm{PCF}(\vec{r})\cdot \textrm{PCF}^{f}(\vec{r})
\end{equation}
and the 2-D PCF enhanced image is given by,
\begin{equation} \label{im_PCF_2d}
g^{\textrm{PCF}}(\vec{r})=\textrm{PCF}^{\textrm{2D}}(\vec{r})\cdot g(\vec{r}). 
\end{equation}
Figs. \ref{pcf_theta_f_map}\subref{a} and \subref{b} show the original PCF enhanced image and the $\textrm{PCF}^f$ enhanced image, respectively. The 2-D PCF enhanced result is demonstrated in Fig. \ref{2d_pcf_im}. 

\section{Simulations and Experimental Results}

This section shows the performance improvement offered by the 2-D CF and PCF in suppressing ghosts and sidelobes, using both simulations and real data. 
\subsection{Simulation results}
\begin{figure}[!t]
	\centering
	\includegraphics[width=2.8in]{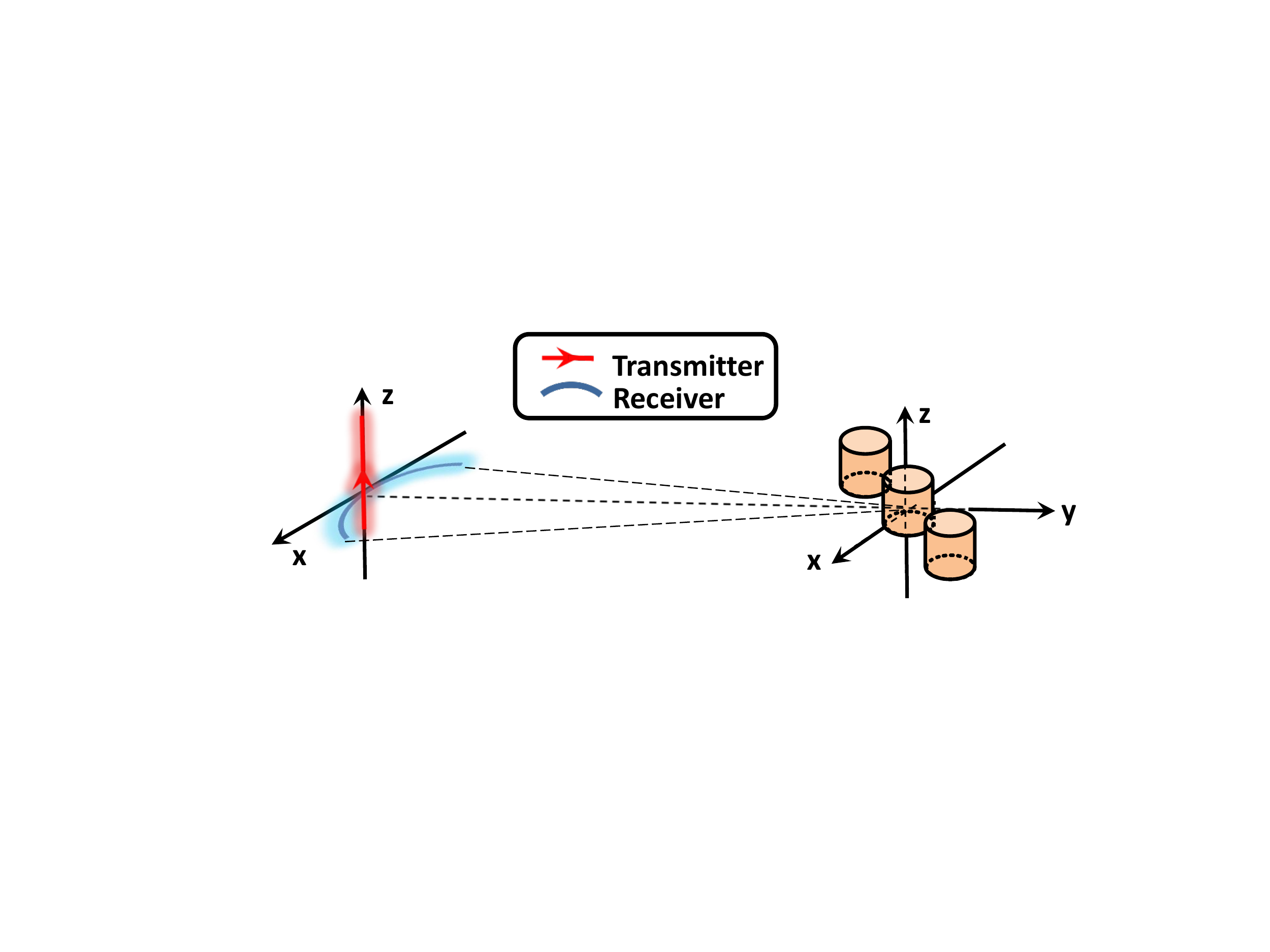}
	\caption{SIMO imaging geometry.}
	\label{simo_model}
\end{figure}

For multipath simulations, we employ ANSYS HFSS, which is a  finite-element method (FEM)-based 3-D electromagnetic field simulator for computing the scattered waves. 
To reduce simulation time, we consider a single input multiple output (SIMO) imaging model, as illustrated in Fig. \ref{simo_model}. There are three targets in the scene. The transmitting antenna is located in the middle of array, whereas the receiving antennas are placed on an arc with a radius of $R_0=10$m. 
The  simulation parameters are given in Table  \ref{tab1}. 


\begin{table}
\centering
\caption{Simulation Parameters for Full Data Set}
\label{tab1}
\setlength{\tabcolsep}{3pt}
\begin{tabular}{|p{100pt}|p{40pt}|}
\hline
Parameters& 
Values \\
\hline
Distance $R_0$&
10 m\\
Start frequency& 
8 GHz \\
Stop frequency&
9 GHz \\
Number of frequency steps&
64 \\
Aperture angle&
$8^{\circ}$ \\
Number of receiving antennas&
81\\

\hline
\end{tabular}
\end{table}

\begin{figure}[!t]
	\centering
	\includegraphics[width=2.5in]{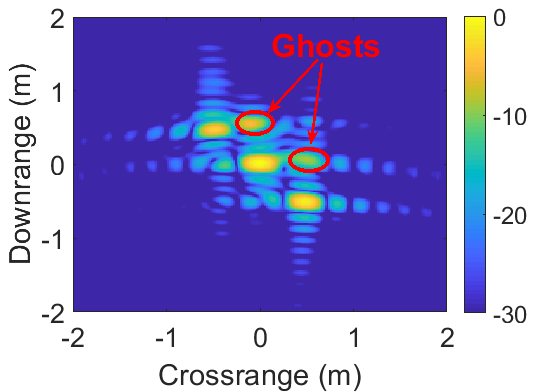}
	\caption{Imaging result by BP algorithm.}
	\label{bp_im_hfss}
\end{figure}

\begin{figure}[!t]
	\centering
	\subfloat[]{
		\includegraphics[width=1.7in]{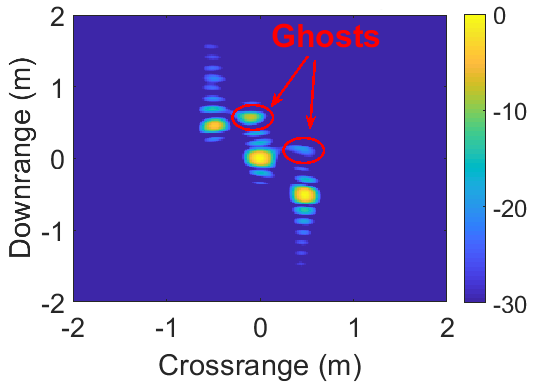}}
	\label{a}\hfill
	\subfloat[]{%
		\includegraphics[width=1.7in]{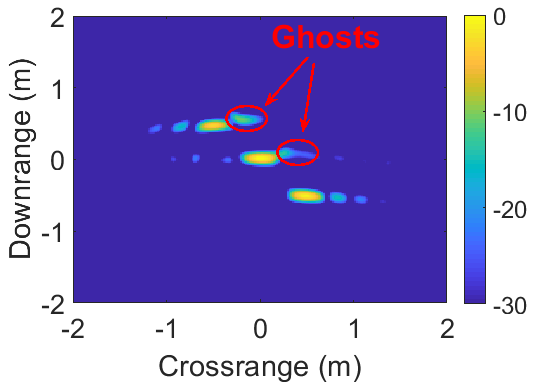}}
	\label{b}\\
	
	\caption{Enhanced images by (a) the original CF,  and (b) the $\textrm{CF}^f$ with incoherent sum in  frequency dimension.}
	\label{cf_theta_f_im_hfss}
\end{figure}

Fig. \ref{bp_im_hfss} shows the imaging result by the BP algorithm. It is  clear that there are two spurious targets labeled as ghosts and shown by red circles. Also, sidelobes are very pronounced both in the range and azimuth dimensions.
The enhanced images by the original CF and the CF$^f$, defined in \eqref{cf_f}, are shown in Figs. \ref{cf_theta_f_im_hfss}\subref{a} and \subref{b}, respectively. 
Note that the ghosts are reasonably suppressed by both factors. However, as evident in Fig. \ref{cf_theta_f_im_hfss}\subref{a}, the original CF is much less effective in suppressing the sidelobes along the range dimension compared to the azimuth dimension. The opposite is true for $\textrm{CF}^f$, as demonstrated in Fig. \ref{cf_theta_f_im_hfss}\subref{b} . 

\begin{figure}[!t]
	\centering
	\includegraphics[width=2.5in]{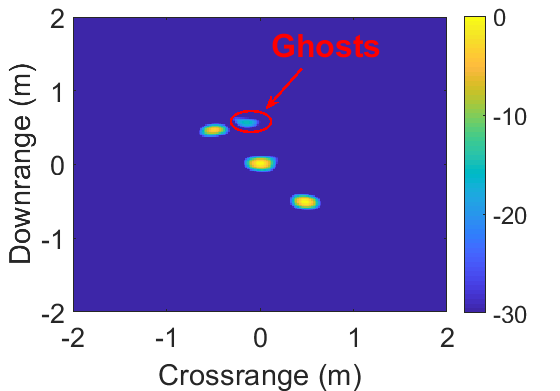}
	\caption{The 2-D CF enhanced image.}
	\label{2d_cf_image_hfss}
\end{figure}

The proposed 2-D CF enhanced image is shown in Fig. \ref{2d_cf_image_hfss}. The sidelobes along the two dimensions are both considerably suppressed. Further, the 2-D CF yields better ghost suppression  by more than 5 dB compared to that offered by the original CF.

\begin{figure}[!t]
	\centering
	\subfloat[]{
		\includegraphics[width=1.7in]{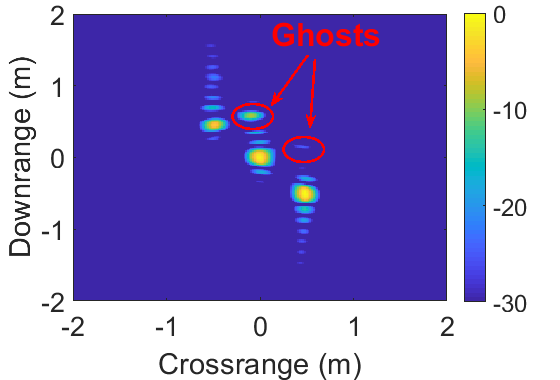}}
	\label{a}\hfill
	\subfloat[]{%
		\includegraphics[width=1.7in]{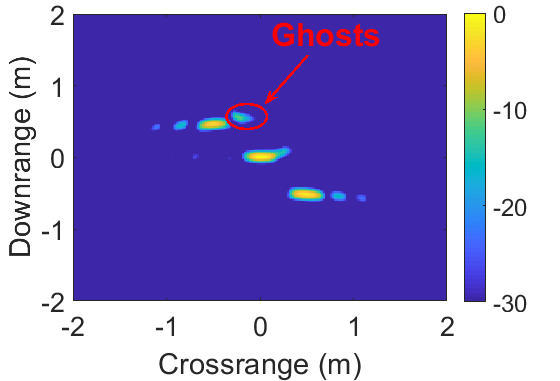}}
	\label{b}\\
	
	\caption{Enhanced images by (a) the original PCF,  and (b) the $\textrm{PCF}^f$ with standard deviation in the frequency dimension.}
	\label{pcf_theta_f_im1}
\end{figure}

\begin{figure}[!t]
	\centering
	\includegraphics[width=2.5in]{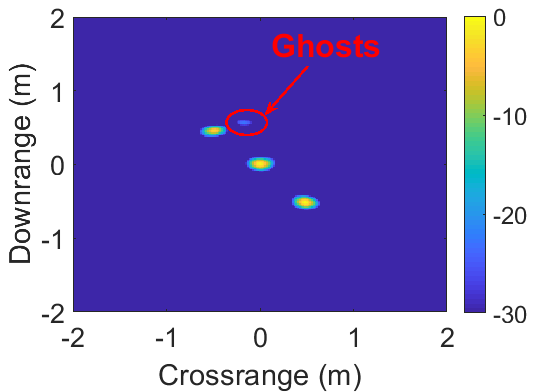}
	\caption{The 2-D PCF enhanced image.}
	\label{2d_pcf_image_hfss}
\end{figure}

Figs. \ref{pcf_theta_f_im1} and \ref{2d_pcf_image_hfss} demonstrate the imaging results by the original PCF, the proposed PCF with standard deviation in the frequency dimension, and the 2-D PCF. The same relative improvement in image quality is evident. 
Comparing Figs. \ref{2d_cf_image_hfss} and \ref{2d_pcf_image_hfss}, one can observe that the PCF based approaches perform slightly better than the CF related approaches.


\subsection{Experimental results}
\begin{figure}[!t]
	\centering
	\subfloat[]{
		\includegraphics[width=1.7in]{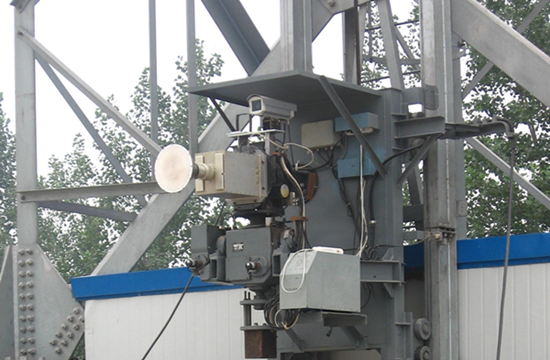}}
	\label{a}\hfill
	\subfloat[]{%
		\includegraphics[width=1.7in]{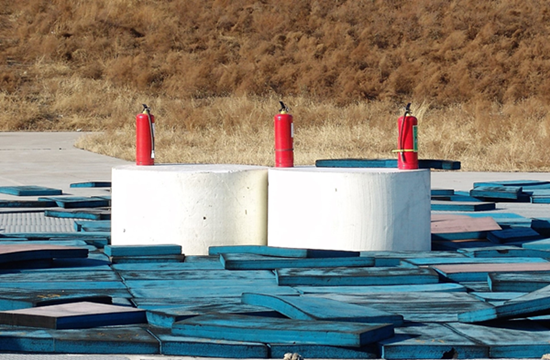}}
	\label{b}\\
	
	\caption{The experimental setup. (a) The measurement radar and (b) target under test: three fire extinguishers.}
	\label{radar_target}
\end{figure}

In addition to simulations, we use an X-band radar to verify the performance of the proposed method. The radar was fixed on a support frame and the target under test was put on a turntable, as shown in Fig. \ref{radar_target}. 

\begin{figure}[!t]
	\centering
	\includegraphics[width=2.5in]{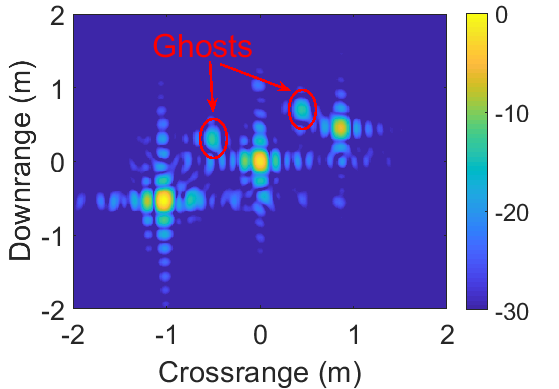}
	\caption{Imaging result by BP algorithm.}
	\label{bp_im_meas}
\end{figure}

\begin{figure}[!t]
	\centering
	\includegraphics[width=2.5in]{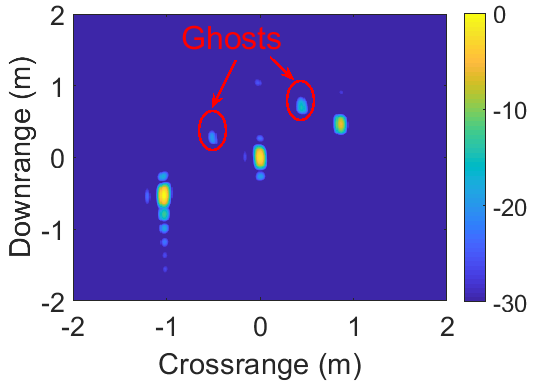}
	\caption{The original CF enhanced image.}
	\label{cf_theta_im_meas}
\end{figure}

\begin{figure}[!t]
	\centering
	\includegraphics[width=2.5in]{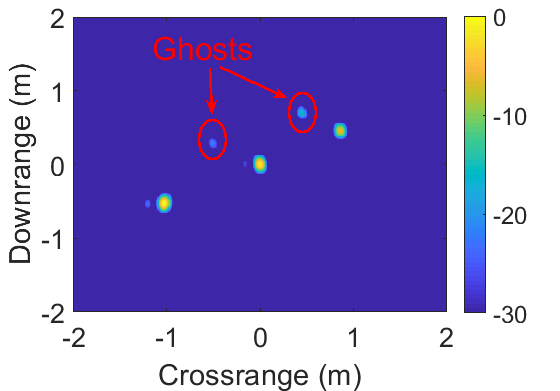}
	\caption{The 2-D CF enhanced image.}
	\label{2d_cf_im_meas}
\end{figure}

Fig. \ref{bp_im_meas} shows the imaging results by the BP algorithm. The ghosts induced by multiple target-to-target reflections are labeled by red circles in the figure. 
The enhanced image by the original CF is given in Fig. \ref{cf_theta_im_meas}, and that of the 2-D CF is shown in Fig. \ref{2d_cf_im_meas}. In the latter, the two dimensional sidelobes, as well as ghosts, are clearly suppressed. 
Here, we do not present the intermediate results of $\textrm{CF}^f$.

\begin{figure}[!t]
	\centering
	\includegraphics[width=2.5in]{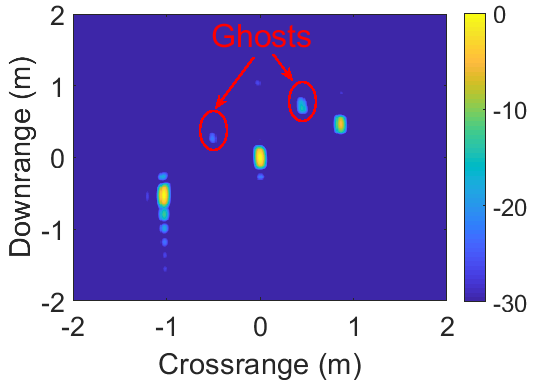}
	\caption{The original PCF enhanced image.}
	\label{pcf_theta_im_meas}
\end{figure}

\begin{figure}[!t]
	\centering
	\includegraphics[width=2.5in]{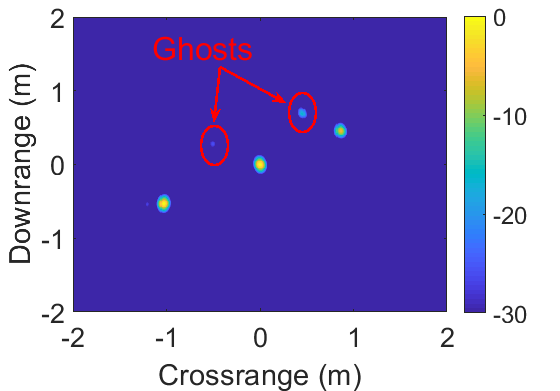}
	\caption{The 2-D PCF enhanced image.}
	\label{2d_pcf_im_meas}
\end{figure}

Figs. \ref{pcf_theta_im_meas} and \ref{2d_pcf_im_meas} show the corresponding imaging results by the original PCF and the proposed 2-D PCF. 

From the above simulations and the real data experiments, the 2-D CF and 2-D PCF lead to higher imaging quality compared to 1-D based suppression techniques.

\section{Conclusions}


This communication proposed a 2-D coherence factor (including phase coherence factor) filtering approach for near-field radar imaging through introducing another incoherent summation in the frequency dimension. 
The incoherent summation  corresponds to diversity of the target spatial spectrum. Due to the use of 2-D diversities, i.e., the antenna diversity and the frequency diversity, the 2-D CF can provide more suppression of sidelobes and ghosts compared with the original versions. 
Simulations and experimental results demonstrated that the proposed approach improves over the original CF or PCF when considering the suppression of sidelobes and ghosts. 

\ifCLASSOPTIONcaptionsoff
  \newpage
\fi

\bibliographystyle{./bibtex/bib/IEEEtran}
\bibliography{./bibtex/bib/full}

\end{document}